\documentclass[usenatbib]{emulateapj}
\usepackage{graphicx}
\usepackage{pslatex}
\usepackage{times}
\usepackage{color}
\usepackage{natbib}

\def\lesssim{\lower.5ex\hbox{$\; \buildrel < \over \sim \;$}}
\def\gtrsim{\lower.5ex\hbox{$\; \buildrel > \over \sim \;$}}
\newcommand{\mstar}{\mbox{$M_{\rm star}$}}
\newcommand{\msun}{\mbox{$M_\odot$}}
\newcommand{\Zsun}{\mbox{$Z_\odot$}}
\newcommand{\fesc}{\mbox{$f_{\rm esc}$}}

\definecolor{gray}{rgb}{0.5,0.5,0.5}

\begin{document}

\title{Heavy dust obscuration of $z=7$ galaxies in a cosmological hydrodynamic simulation}

\author{Taysun Kimm \& Renyue Cen}
\affil{Department of Astrophysical Sciences, Princeton University, Peyton Hall, Princeton, NJ 08544, USA}

\shorttitle{UV and optical properties of $z=7$ galaxies}
\shortauthors{Kimm \& Cen}

\begin{abstract}
Hubble Space Telescope observations with Wide Field Camera 3/IR reveal that galaxies at z$\sim$7 have 
very blue ultraviolet (UV) colors, consistent with these systems being dominated by young stellar 
populations with moderate or little attenuation by dust. We investigate UV and optical properties of the high-$z$
galaxies in the standard cold dark matter model using a high-resolution adaptive mesh refinement 
cosmological hydrodynamic simulation. For this purpose, we perform panchromatic three-dimensional 
dust radiative transfer calculations on 198 galaxies of stellar mass $5\times10^8-3\times10^{10}$ \msun\ 
with three parameters, the dust-to-metal ratio, the extinction curve, 
and the fraction of directly escaped light from stars (\fesc). Our stellar mass function is found to be in broad 
agreement with Gonzalez et al., independent of these parameters. We find that 
our heavily dust-attenuated galaxies ($A_V\sim1.8$) can also reasonably match modest UV--optical colors, blue UV slopes, 
as well as UV luminosity functions, provided that a significant fraction ($\sim$10\%) of light directly escapes from them.
The observed UV slope and scatter are better explained with an Small Magellanic Cloud-type extinction curve, 
whereas Milky Way-type dust predicts too blue UV colors due to the 2175 $\AA$ bump.
We expect that upcoming observations by ALMA will be able to test this heavily obscured model.
\end{abstract}

\keywords{galaxies: high-redshift --- galaxies: ISM}

\section{Introduction}
Observational techniques based on photometric properties of star-forming galaxies 
have proven to be successful in identifying galaxies at high redshift \citep{steidel96}.
In particular, the arrival of recent Wide Field Camera 3 / Infrared (WFC3/IR) data taken 
with the Hubble Space Telescope has enlarged the sample of $z$-dropout galaxies 
($z\sim7$) \citep[e.g.,][]{bouwens12a}, allowing for the investigation of ultraviolet (UV) 
properties of the early galaxies. These $z\sim7$ candidates reveal $(FUV-NUV)$ colors 
close to zero or sometimes negative 
\citep{bouwens10,wilkins11,mclure11,dunlop12,finkelstein12,bouwens12a,dunlop13,bouwens13}, 
indicating a very blue spectral energy distribution with $-2.5\lesssim\beta\lesssim-1.5$, 
where the UV-continuum slope $\beta$ relates to the flux density per unit wavelength ($f_\lambda$) 
as $f_\lambda\propto\lambda^\beta$. Given that a small amount of dust can substantially redistribute 
the spectral energy distribution, the low $\beta$ value has been taken 
to indicate a limited amount of dust in these systems. 
 
On the other hand, using $Spitzer$ Infrared Array Camera (IRAC) and WFC3/IR data, \citet{labbe10} reported 
that their $z\sim7$ galaxies have moderate rest-frame optical colors $U-V\sim0.3-0.5$,
implying the presence of evolved stellar population older than $100$ Myr. 
In fact, for a realistic rising star formation (SF) history, the age of the population alone seemed 
very difficult to explain such colors without significant attenuation by dust and/or nebular line 
emission \citep{finlator11,labbe10}. Several authors pointed out the possibility that 
these apparently conflicting results are may be due to contamination by nebular emission lines, 
such as H$\beta$ and [O{\sc iii}] 4959, 5007 doublet 
\citep{schaerer10,labbe12,curtis-lake13,stark13,schenker13,bouwens13}, to the IRAC 3.6 $\mu$m band. 
For example, \citet{labbe12} claim that the $z\sim7$ galaxies show $H_{160}$--$[4.5]$ colors that are much bluer 
than $H_{160}$--$[3.6]$ colors, supporting the view that the high-$z$ galaxies are young and affected by moderate 
or little dust attenuation.

However, some hydrodynamic simulations in a cold dark matter cosmology predict that 
galactic metal enrichment occurs in a very short time scale \citep[e.g.,][]{finlator11},
leading to the formation of galaxies of stellar mass $10^{10}\,\msun$ with a metallicity close to solar at $z\sim7$.
Given that galaxies are more compact and denser at higher redshift \citep[e.g.,][]{ferguson04,trujillo06,van-dokkum08},
one may expect that the simulated galaxies are significantly dust attenuated,
and may be in conflict with observations. In this study, we test this idea by 
confronting simulated UV luminosity and stellar mass functions, UV and optical colors of $z=7$ galaxies 
with observations. A self-consistent three-dimensional radiative transfer calculation on galaxies produced 
by a cosmological simulation with $\sim$ 30 times better spatial resolution (29 $h^{-1}$ pc) is a significant 
improvement over previous studies \citep{devriendt10,salvaterra11,finlator11}.
More importantly, we show that, with an Small Magellanic Cloud (SMC)-type extinction curve, 
our substantially dust attenuated galaxies can also simultaneously explain the observed red 
UV--optical colors, blue UV colors, as well as UV luminosity functions if the escape fraction of stellar 
photons is $\sim$10\%. 

The outline of this paper is as follows. We first describe the simulation details in Section 2. 
In Section 3, simulated galaxies with different 
assumptions on dust properties are compared with observations. Finally, we discuss  
implications of these results and conclude in Section 4.

\begin{figure*}
  \centering
      \includegraphics[width=8.6cm]{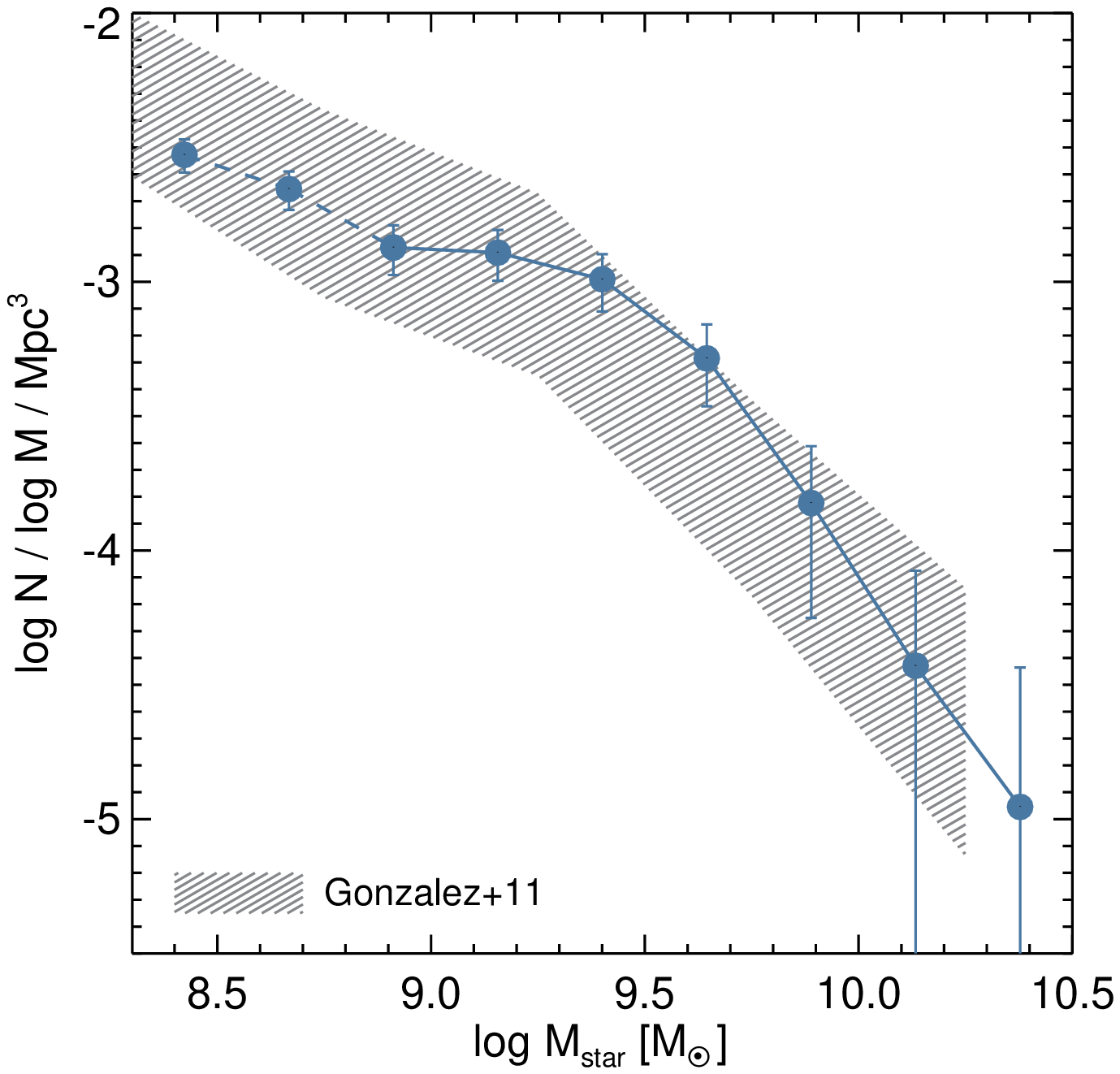}
      \includegraphics[width=8.6cm]{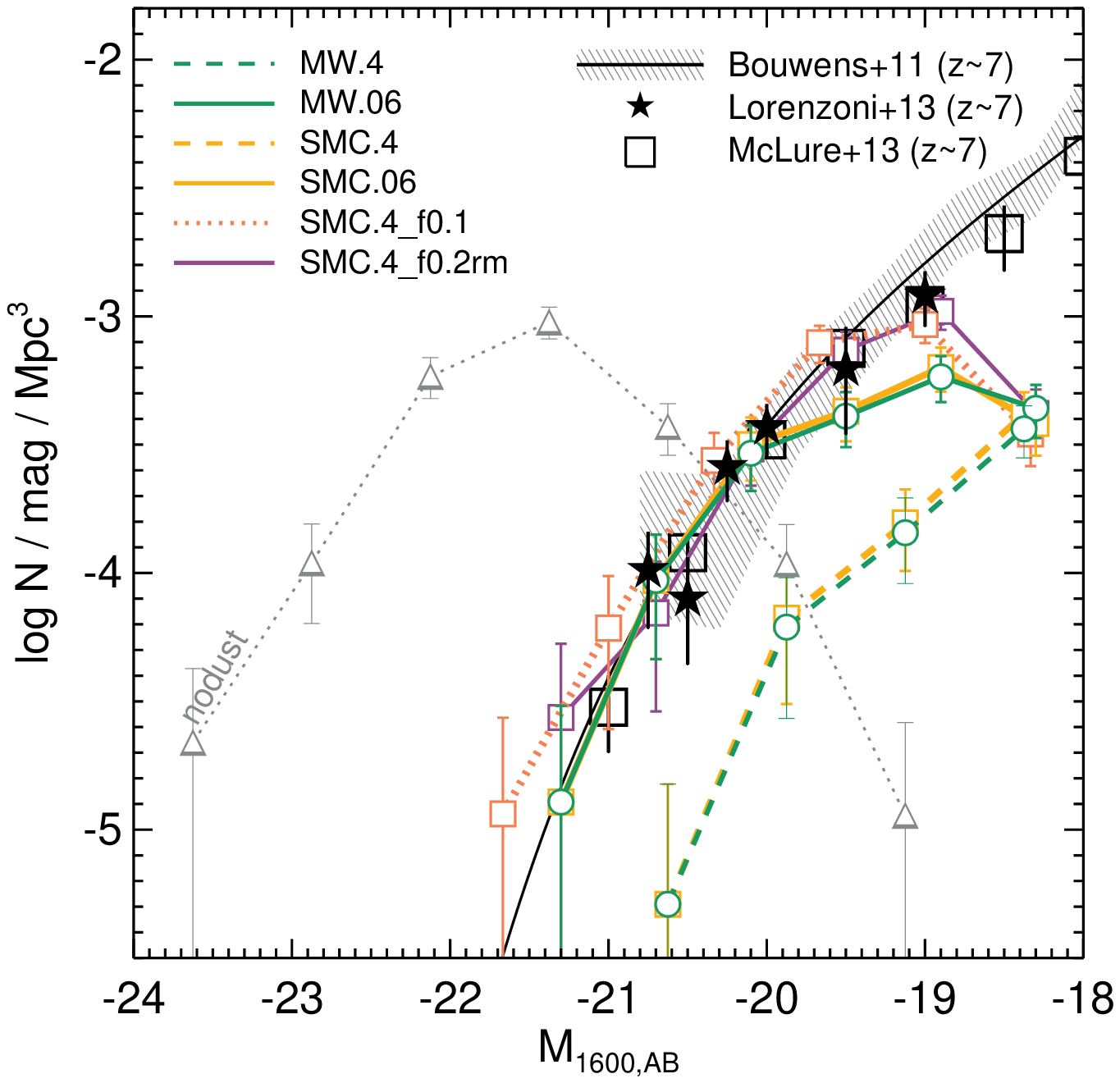} 
 \caption{
Left panel: stellar mass function (SMF) at $z=7$. 
 Our SMF shows a reasonable match to the observed SMF \citep[][gray shading]{gonzalez11}.
 In this work, we only make use of galaxies with $\mstar \ge 5\times10^{8}\,\msun$ to minimize the effect of 
finite resolution (solid line).
Right panel: Rest-frame UVLFs at $z=7$ with different assumptions on dust properties. 
 The UVLF without dust attenuation is shown as a gray dotted curve with open triangles. 
Note that the peak and downturn at $M_{1600}>-21$ are due to resolution effects.
 Green solid and dashed curves with open circles are the UVLFs obtained using the Milky Way (MW)-type extinction curve 
 with a dust-to-metal ratio ($D/M$) of 0.06 and 0.4, respectively;
the corresponding orange curves with open squares are computed using SMC-type dust. 
Error bars denote 1$\sigma$ Poisson. Observed UVLFs by \citet{bouwens11a}, \citet{lorenzoni13}, 
and \citet{dunlop13} are shown as gray shading, black stars, and black squares, respectively.
A dotted orange line (${\rm SMC.4\_f0.1}$) corresponds to the case in which 10\% of light from 
stars is assumed to directly escape from galaxies. 
We also present an {\em ad hoc} case with $\fesc=0.2$ and $D/M=0.4$
in which stellar mass and the amount of metals are reduced by a factor 2.5 (purple solid line, see text).}
\label{fig:lf}
\end{figure*}

\section{Simulations}

The cosmological simulation is performed with the Eulerian hydrodynamics code, {\sc enzo} \citep[][]{1999aBryan, 2005OShea, 2009Joung,bryan13}.
For more details on the simulation setup and implemented physics, the reader is referred to \citet[][]{2012Cen}.
Briefly, this is a large zoom-in simulation of $21\times24\times20\,h^{-3}\,{\rm Mpc^3}$  
embedded in a $120 \,h^{-1}\, {\rm Mpc} $ periodic box.
The simulation includes a uniform Haardt-Madau UV background where the cosmic reionization occurs at $z=9$, 
a shielding of UV radiation by neutral hydrogen, metal-dependent radiative cooling down to $10$K, 
star formation, and supernova feedback \citep[][]{2012Cen}.
The physical prescriptions have been validated by a large suite of independent observations, including 
the cosmic star formation history \citep{cen11}, 
and damped Lyman alpha systems at redshift $z=0-5$ \citep{2012Cen}. 
The maximum spatial resolution is $29\,h^{-1}\,{\rm pc}$, and dark matter particles
of $1.3\times10^7\, h^{-1}\,\msun$ are used within the zoomed-in region. 
The initial condition is generated with the cosmological parameters that are consistent 
with the WMAP7 results \citep{komatsu11}:
$(\Omega_{\rm m}, \Omega_{\rm b}, h, \sigma_8, n_s  = 0.28, 0.046, 0.72, 0.82, 0.96)$.

To compute the spectral energy distributions (SEDs) of each galaxy, 
we post-process the simulation output at $z=7$ using a three-dimensional dust radiation transfer code, 
{\sc sunrise} \citep{jonsson06,jonsson10}. The main advantage of {\sc sunrise} is the use of a 
polychromatic algorithm, which can trace information in all wavelengths per each ray.
It makes use of the standard dust cross-sections \citep[e.g.,][]{weingartner01,draine07} to 
simulate absorption and multiple scattering by dust. 
The input stellar spectrum is taken from {\sc starburst99} \citep{leitherer99} 
assuming a Kroupa initial mass function with the low (high) mass cut-off of 0.1 \msun\ (100 \msun).
{\sc sunrise} also uses the spectrum of {\sc Hii} or photo-dissociation regions (PDRs) computed by a 
photo-ionization code, {\sc mappingsiii} \citep{dopita05,groves08}, to take into account the 
immediate absorption and emission by birth clouds. This is done by replacing SEDs of 
young ($\le 10\,{\rm Myr}$) star particles with re-processed SEDs of a population with 
constant star formation for 10 Myr by {\sc mappingsiii} \citep[see][]{jonsson10}. 
The fraction of light processed by PDRs is controlled by a 
parameter, $f_{\rm PDR}$, which we adopt $f_{\rm PDR}=0.2$ following \citet{jonsson10}.
The amount of dust is inferred from the amount of metals 
by assuming an adjustable dust-to-metal ratio ($D/M$).
We use the maximum resolution available from the AMR hydrodynamic 
simulation ($\sim 29 \, h^{-1}$) to compute absorption and scattering by dust.  
We test the convergence of the Sunrise results by degrading the most 
refined cells in each halo by one level, before performing dust radiative transfer.
We find that the attenuated UV and UV-optical colors are well converged 
(within $\sim$ 0.01 dex).
Note that the attenuated spectra is redshifted to $z=6.7$ in order to match the typical redshift of the $z\sim7$ sample \citep[e.g.,][]{bouwens13} 
and then convolved with WFC3/IR F125W, F160W 
and IRAC  $[3.6\mu m]$  filter throughputs to yield $J_{125}$, $H_{125}$, and $[3.6]$ mag in AB magnitude, 
respectively.

\section{Results}

Our simulated sample consists of 198 galaxies of stellar mass $5\times10^8 \le \mstar \le 2.5\times10^{10}\,\msun$ 
at $z=7$. The simulated galaxies with  $\mstar \simeq 5\times 10^9\,\msun$ have the specific star formation rate 
(sSFR) of $\simeq 6 \,{\rm Gyr^{-1}}$, with more massive galaxies having smaller sSFR \citep[see also][]{cen11},
compatible with the recent measurements by \citet{schaerer10,bouwens12a,stark13}.
A simple fit to the stellar mass and SFR gives 
$ {\dot M}_{\rm star} \simeq 55 \,\msun\ {\rm yr^{-1}} \, \left(\mstar/10^{10}\msun\right)^{0.76} $. 
Predicted stellar (gas) metallicities of galaxies with $\mstar \simeq 10^9$, and 
$10^{10}\,\msun$ are $\simeq$ 0.5, and 1.0\,\Zsun\ (0.6, and 0.5 \Zsun), respectively, 
indicating that more massive galaxies have more metal-rich populations.
We note that similar metal enrichment is found in \citet[][Fig.~11]{finlator11}, 
which is based on an entirely independent numerical technique and feedback prescription.

\subsection{Calibrations to stellar mass and UV luminosity functions}
\label{sec:uvlf}

The left panel of Fig.~\ref{fig:lf} shows the stellar mass function (SMF) of simulated galaxies at $z=7$,
which is seen to match reasonably well the observed SMF based on SED fitting with $HST$+$Spitzer$ 
data \citep{gonzalez11,labbe10}. 
However, we note that nebular emission lines, such as $H\beta$ or {\sc [Oiii]} 5007, may
contribute to Spitzer/IRAC 3.6 $\mu$m bands, resulting in possible overestimation of 
stellar mass of observed galaxies \citep{schaerer10,stark13,curtis-lake13}. 
Bearing this in mind, we also discuss a model in which stellar mass, stellar and gas metallicities 
are artificially lowered by a factor of 2.5, such that galaxies with  $\mstar \simeq 10^9\msun$ have 
stellar metallicity of $\simeq 0.2\,\Zsun$.

In the right panel of Fig.~\ref{fig:lf}, we present rest-frame UV luminosity functions (UVLFs) at $z=7$. 
The absolute UV magnitude ($M_{1600}$) is measured by integrating the SED at $\lambda=1600\,\AA$ 
through a square filter of width $100\,\AA$. We find that the {\it intrinsic} UV of simulated 
galaxies (gray dotted line) would be too bright compared with observations   
\citep{bouwens11a,lorenzoni13,dunlop13} by about 2--3 magnitudes.
Performing dust radiative transfer substantially improves the agreement of simulated 
UVLFs with observations. We find that a dust-to-metal ratio of $D/M=0.06$ successfully reproduces 
the observed UVLFs up to $M_{\rm UV} \approx -19.5$ (corresponding to $3-10\times10^{8}\,\msun$, 
e.g., \citealt{stark13}), above which our simulation is incomplete. The resulting attenuation in FUV 
is found to be $1\lesssim A_{\rm 1600} < 5$.
On the other hand, adopting the dust-to-metal ratio $D/M=0.4$, inferred from local metal-rich 
galaxies \citep[e.g.,][]{dwek98,draine07b}, we find that the simulated galaxies would suffer 
from too much extinction (dashed lines). This is the case regardless of whether 
the extinction curve is MW-type (green lines) or SMC-type (orange lines), as the dust extinction 
cross-section around 1600 $\AA$ is similar in the two cases \citep[e.g.,][]{weingartner01}.

We stress that it is possible that our simulation under-resolves the porosity of the interstellar medium, 
potentially giving rise to a higher covering fraction of dusty gas around young stars \citep[c.f.,][]{wise09}.
In this case, the number of UV photons directly coming out of galaxies may be under-estimated.
We explore this possibility by introducing a free parameter, $f_{\rm esc}$, which quantifies 
the fraction of directly escaped light from stars.  Conceptually,  $f_{\rm esc}$ differs from 
$f_{\rm PDR}$ in the sense that the former is related to the global ISM structure, 
while the latter is only applicable to young stars and their birth clouds.
We find that adding 10\% of the intrinsic stellar light to the strongly attenuated spectrum 
with $D/M=0.4$ can also give a reasonable match to the observed UVLF (orange dotted line), 
indicating that there is a degeneracy between a dust-to-metal ratio and the escape fraction. 
This particular case may be viewed as the model with the maximum $\fesc$, 
as opposed to the models with $\fesc=0$ (MW.06 or SMC.06).
The resulting difference between the two cases is the differential reddening in UV and optical bands, 
as we discuss in the next section.

It is worth noting that our simulated galaxies represent actively star-forming, dusty galaxies.
In the case of the model with SMC-type extinction, $D/M=0.4$, and $\fesc=0.1$ (SMC.4\_f0.1),
 galaxies with $5\times10^{8} \le \mstar < 3\times 10^{10}\,\msun$ turn out to have 
$A_V \simeq 1.8\pm 0.3$.
Given that high-$z$ galaxies are generally more compact than the local counterpart \citep[e.g.,][]{ferguson04}, 
such heavy attenuation may not be too surprising.
As aforementioned, however, if our galaxies formed too many stars and overproduced metals,
the extinction is likely to be overestimated. To quantify a possible change in the extinction, 
we compare an {\em ad hoc} case (${\rm SMC.4\_f0.2rm}$) in which stellar mass, metallicity, and gas metallicity 
are reduced by a factor of 2.5 before performing radiative transfer calculations. We use an 
escape fraction of $20\%$ to match the observed UVLFs in this case.
Even in this model with smaller galaxy masses and smaller amounts of dust, it turns out that the galaxies  
still show more significant attenuation of $\left<A_V\right> \simeq1.4$  
compared to $\left<A_V\right> \sim 1$ derived based on a single dust screen model fitting of the 
observed SED \citep{schaerer10}.

\begin{figure}
  \centering
      \includegraphics[width=8.6cm]{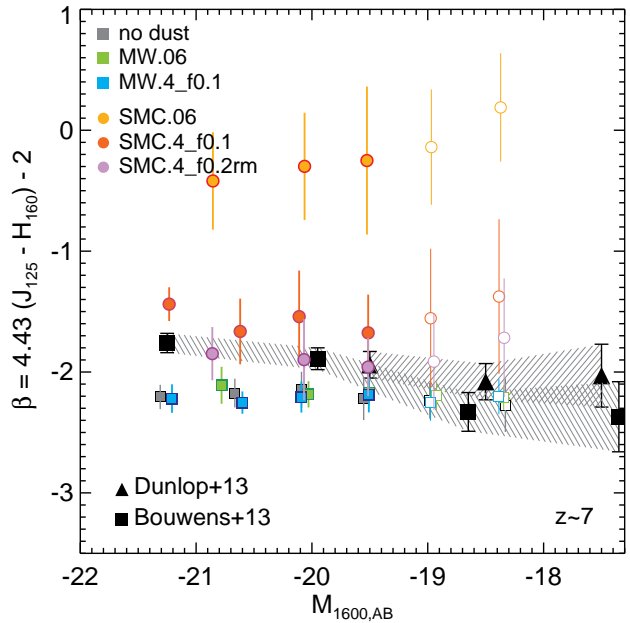} 
 \caption{Comparison of UV-continuum slopes between observations and models at $z\sim7$. 
 Different colors and symbols correspond to UV slopes obtained with different dust properties and escape fractions, as indicated in the legend. 
 Two latest observational determinations by \citet{bouwens13} and \citet{dunlop13} are shown 
 as filled squares and triangles with the standard error in the mean, respectively.  Gray squares denote the intrinsic UV slope and intrinsic UV luminosity 
 shifted by two magnitudes (i.e., $M_{\rm UV}^{\rm int}+2$). We display less reliable estimates due to 
 incomplete galaxy distributions (see Fig.~\ref{fig:lf}) with empty symbols. 
 Each bin contains at least three galaxies, and the error bars show the standard deviation.
 Predicted UV slopes with MW-type dust turn out to be too steep in the bright galaxies,
whereas SMC-type dust with a high escape fraction predicts slopes that are consistent, within error bars, with observations.
 }
\label{fig:slope}
\end{figure}

\subsection{UV and optical colors}
\label{sec:colour}

Fig.~\ref{fig:slope} shows the UV slope ($\beta$) as a function of 
UV luminosity, where $\beta$ is computed from the UV color as $\beta = 4.43  (J_{125}-H_{160}) - 2$ \citep{dunlop13}.
The latest determinations of the slope from two independent studies, \citet{bouwens13} and \citet{dunlop13}, 
are shown as squares and triangles. Several interesting features can be found in this figure. 
First, our simulated galaxies show intrinsically blue UV colors ($\beta\sim-2.2$) 
with little dependence on luminosity. Although SFRs of our simulated galaxies increase 
with increasing stellar mass, we find that a considerable scatter in the $M_{\rm star}-\dot{M}_{\rm star}$ 
relation ($\sim 0.3$ dex) leads to the scatter in the intrinsic $M_{\rm UV}$ for a given stellar mass, 
weakening the dependence on the UV slope (Fig.~\ref{fig:slope}, gray squares). 
Second, the UV slope in the models with MW-type dust turns out to be too steep for bright galaxies ($M_{UV} \lesssim -20$)
regardless of the amount of dust (MW.4\_f0.1 and MW.06). 
This is because the dust cross-sections in $J_{125}$ and $H_{160}$ bands are comparable due to the 2175 $\AA$ bump.
For the same reason, the models predict a much smaller scatter in the UV slope than observed.
These suggest that dust in the $z\sim7$ galaxies is unlikely to share the same grain distribution as Milky Way. 
Third, the model in which SMC-type dust pervades (SMC.06) predicts too shallow UV slope ($\beta\sim-0.5$) 
than observed due to stronger attenuation at FUV than at NUV wavelengths.
We find that predicted UV slopes of luminous galaxies ($M_{UV} \lesssim -19.5$) are broadly consistent with observations, 
within error bars, only in the models with SMC-type dust and a large escape fraction (SMC.4\_f0.1 and SMC.4\_f0.2rm).
For the less luminous bins ($M_{UV}\gtrsim -19.5$), 
our simulation is incomplete (Fig.~\ref{fig:lf}), and thus may not be compared directly to the observations.
Nevertheless, the correlation between the UV slope and luminosity in this model appears to be weak 
in the bright regime, consistent with observations. 
It is also noteworthy that the scatter in the UV slope is more notable than that of MW-type dust,
as there is a mixture of galaxies with different attenuation for a given UV luminosity.

Another factor that can alter the UV slope is the covering fraction of the PDR ($f_{\rm PDR}$). 
As shown in Fig.~6 from \citet{groves08}, a covering fraction close to 1 
considerably reddens the UV spectrum of young stellar populations. 
We examine how sensitive the predicted UV slope is to the choice 
of $f_{\rm PDR}$, and find that our results do not depend strongly on the parameter. 
In the case of $f_{\rm PDR}=0$, the UV slope of the input SED (i.e. {\sc mappingsiii} + {\sc starburst99} ) 
is only slightly steeper ($\Delta \beta = -0.045\pm0.034$) than the one with the fiducial value, $f_{\rm PDR}=0.2$.
Conversely, setting $f_{\rm PDR}=0.5$ makes the slope shallower by only +0.09 dex. After extinction, 
the difference becomes even smaller. Such a small change in $\beta$ is essentially because i) the 
impact of $f_{\rm PDR}$ on the UV slope is only prominent at $f_{\rm PDR}\gtrsim0.5$, 
and ii) a significant fraction ($\sim 40\%$) of the intrinsic UV flux arises from the stars older than 10 Myr. 
However, even when $f_{\rm PDR}=1$ is assumed, the change in the slope of the input SED is found to be 
$\Delta \beta\sim0.4$, which is a smaller effect than that of attenuation by the intervening interstellar dust 
or the escape fraction.

\begin{figure}
  \centering
      \includegraphics[width=8.6cm]{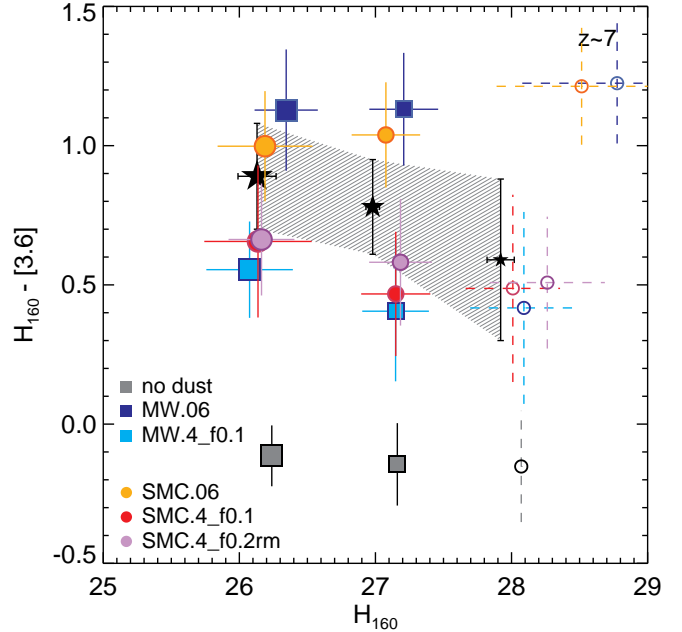} 
 \caption{Comparison of UV--optical colors of galaxies at $z\sim7$ between observations 
 \citep[][black stars]{labbe10} and models with different assumptions on dust properties (color circles).
  For a fair comparison with \citet{labbe10}, we present average colors and their dispersions 
 for three $H_{160}$ magnitude bins, $25 < H_{160} \le 26.6$, 
$26.6 < H_{160} \le 27.5$,  and $H_{160} \ge 27.5$, except for the case without dust (gray circles). 
The last bin is incomplete due to finite resolution, and shown as open circles.
 We add the contribution from nebular line emission corresponding to the 
 equivalent width of 320 $\AA$ to the [3.6] flux (0.3 dex in mag) in the  dust attenuated cases,
 motivated by observations. Our galaxies are intrinsically blue in UV--optical colors (gray), 
 while the models with dust attenuation could produce UV--optical colors that are compatible with observations.
 }
\label{fig:colour}
\end{figure}

An interesting puzzle of $z\sim7$ galaxies is their red UV--optical color  $H_{160}$--$[3.6])\sim0.6 - 0.9$. 
As pointed out by \citet{finlator11}, obtaining such a red UV--optical color is not trivial for actively 
star-forming galaxies with little dust.
In accord with \citet{finlator11}, we predict that our simulated galaxies are indeed very blue, 
$(H_{160} - [3.6])\sim -0.1$, in the absence of dust attenuation and nebular emission lines, shown as 
gray filled circles in Fig.~\ref{fig:colour}.
For a fair comparison with \citet{labbe10}, we present average colors and their $1\sigma$ dispersions
by dividing the simulated galaxy sample into three $H_{160}$ magnitude bins, $25 < H_{160} \le 26.6$, 
$26.6 < H_{160} \le 27.5$,  and $H_{160} \ge 27.5$. 
Notice that the last bin ($M_{\rm UV} \gtrsim -19.5$) is incomplete due to finite resolution (Fig.~\ref{fig:lf}), 
and displayed as empty circles. 

Inclusion of nebular emission lines can improve the agreement 
by increasing the flux in the optical band. Strictly speaking, several strong emission lines, 
such as {\sc [Oii] 3726/3729} doublet or {\sc [Oiii] 4959, 5007}, do not fall within 
{\it Spitzer}/IRAC $[3.6]$ window for $z=7$ galaxies. However, for galaxies at redshift $z=6.7$,  
{\sc [Oiii] 4959, 5007} lines can contribute to the $[3.6]$ flux. In order to take into account the 
contribution, we add the emission lines corresponding to the equivalent width of 
$\simeq 320\,\AA$, motivated by observations\footnote{We could directly compute the equivalent 
width of the emission lines from the {\sc sunrise} output, but we decided not to do so, mainly 
because the prediction is uncertain in the case of the models with a non-zero escape fraction. 
Accordingly, we removed the emission lines from the simulated SEDs, and then increased the $[3.6]$ 
magnitude by 0.3 dex, motivated by observations \citep{labbe12,schenker13}.}\citep{labbe12,schenker13}. 
This means that the UV bright (faint) galaxies are likely to have moderate (blue) UV--optical {\it continuum} colors.
Nevertheless, inclusion of nebular emission does not suffice to account for the observed colors, 
indicating that it may not be fully responsible for the observed red $H_{160}$--$[3.6]$ colors 
\citep[see also][c.f. \citealt{stark13}]{finlator11}. 

We find that red UV--optical colors can naturally arise by dust extinction.
Fig.~\ref{fig:colour} clearly shows that the models with a zero escape fraction (MW.06 and SMC.06) 
produce sufficiently red galaxies $(H_{160} - [3.6] \gtrsim 1)$,
as the dust extinction cross-section is larger in the UV wavelengths than in the optical range \citep{weingartner01}.
In particular, the models showing reasonable UV slopes (SMC.4\_f0.1 and SMC.4\_f0.2rm) predict UV--optical colors that are 
compatible with observations. Note that the contribution from nebular line emission (0.3 dex) is already 
included in theses calculations, and dust extinction accounts for the reddening 
of $\sim 0.4$ dex in the two models (SMC.4\_f0.1 and SMC.4\_f0.2rm).
We expect that higher dark matter resolution would lead to better agreement,
as more stars would form in smaller dark matter halos at earlier epochs.
For example, we find that increasing the mean stellar age from 100 Myr to 200 Myr can easily redden 
the UV--optical colors by 0.2 dex without changing UV slopes significantly. 
Thus, we conclude that our heavily obscured galaxies with SMC-type dust and a high escape fraction 
can also simultaneously explain the blue UV colors and moderate UV-optical continuum colors.

\section{Conclusions and Discussion}

Utilizing an ab initio cosmological adaptive mesh refinement hydrodynamic simulation 
with high resolution ($29\, h^{-1}$ pc), we investigate $FUV$--$NUV$ and $UV$-optical colors
of 198 simulated galaxies at $z=7$. We post-process our metal-rich galaxies ($\sim 0.1-1.0\Zsun$)  
through three-dimensional panchromatic dust radiative transfer calculations by varying three parameters, 
the dust-to-metal ratio ($D/M$), the extinction curve, and the fraction of directly escaped light (\fesc). 
By matching simulated UV luminosity functions to the ones observed, 
we identify two sets of models with ($D/M$, \fesc) = (0.06, 0) or (0.4, 0.1), 
corresponding to the cases with the minimum and the maximum escape fraction.

We find that the observed moderate UV--optical continuum colors 
as well as blue UV colors of $z\sim7$ galaxies can be reproduced simultaneously 
with SMC-type dust, a large dust-to-metal ratio of 0.4, and \fesc=10\%.
The resulting attenuation in the V band is found to be more significant ($\left<A_V\right>\sim1.8$) than 
the one derived based on a single dust screen model fitting of the observed SED \citep{schaerer10}.
On the other hand, the model with a zero escape fraction and $D/M=0.06$ 
produces galaxies with too shallow UV slopes ($\beta\sim-0.5$) compared with 
observations \citep{dunlop13,bouwens13}. 

After searching through the parameter space, the fact that the model with a $\fesc\sim10\%$ 
is favored over to a model with much smaller \fesc\ is worth noting.
Observations of cosmological reionization infer the Thomson optical 
depth $\tau_e = 0.089 \pm 0.014$ \citep{hinshaw12}, indicating a reionization
redshift of $z_{re}=10.6 + 1.1$ (assuming a sudden reionization picture). 
In order to reionize the universe at this redshift range by stellar sources, $\fesc\sim10\%$ is 
likely required  \citep[e.g.,][]{cen03}.
Moreover, detailed radiative transfer simulations indicate a porous interstellar medium and,
$f_{\rm esc}\sim10\%$ is within the range of predictions for galaxies at high redshift
\citep{wise09,razoumov10,yajima11}.

By contrast, we find that predicted UV slopes are too steep regardless of the amount of dust 
in the case of MW-type dust due to the presence of the 2175 \AA\ bump.
The bump also results in a much smaller scatter in the UV slope than observed, 
suggesting that $z\sim7$ galaxies are more likely to have SMC-type dust.  
The most successful models (SMC.4\_f0.1 and SMC.4\_f0.2rm) predicts the UV slope to be weakly 
dependent on UV luminosity in the range $-22 < M_{\rm UV} \lesssim -19.5$, 
as shown in observations \citep{bouwens13,dunlop13}. These conclusions remain largely unchanged 
even if we assume that our galaxies overproduced stars and metals by a factor 2.5 
to take into account possible overestimation in stellar mass of observed galaxies \citep{schaerer10,stark13,curtis-lake13},  
except that $\left<A_V\right>$ becomes slightly lower $\sim 1.4$.

Our simulated galaxies are found to be dusty, and heavily affected by attenuation ($1\lesssim A_{\rm FUV} < 5$).
The predicted attenuation is larger than $1\lesssim A_{\rm FUV} \lesssim 3$
in star-forming galaxies with similar masses at lower redshift ($1\le z \le 2.5$) \citep[e.g.,][]{buat12},
but in a direction expected since high-$z$ galaxies are more compact than 
the low-$z$ counterparts \citep[e.g.,][]{ferguson04}. 
An important question is whether a large amount of dust can form in galaxies at $z\sim7$,
given that a substantial fraction of stardust may originate from the late stage of stellar evolution \citep[e.g.,][]{draine09}.
Interestingly, observations of quasars at $z\sim6-7$ report the detection of large masses of 
dust \citep{bertoldi03,dwek07,wang08,venemans12}, supporting that a majority of dust is 
grown in the ISM \citep[][see also for an alternative explanation by \citealt{bianchi07}]{draine03}.
A most direct test of our model is by detection of strong infrared emission from 
the UV selected galaxies and details will be presented in a companion 
paper (Cen \& Kimm 2013, {\sl in prep.}).

\section*{Acknowledgements}
We are grateful to the referee for a careful review of the paper.
We also thank Silvio Lorenzoni and Rychard Bouwens for sharing their data, 
and the YT community \citep{turk11} for providing useful analysis routines. 
Computing resources were in part provided by the NASA High-
End Computing (HEC) Program through the NASA Advanced
Supercomputing (NAS) Division at Ames Research Center.
The research is supported in part by NSF grant AST-1108700 and NASA grant NNX12AF91G.

\small
\bibliographystyle{apj}
\bibliography{ms}

\end{document}